\documentclass{article}
\usepackage{graphicx}
\usepackage{cite}
\begin{document}
\renewcommand{\thefootnote}{\fnsymbol{footnote}}
\begin{center}
\section*{\Large{Spontaneous formation of helical states in polymer chains via molecular dynamics simulation }}

{\large{S.A.Sabeur$^{a,*}$, F.Hamdache$^a$, M.Bouarkat$^b$}}

 {(a) \it Laboratoire de Physique des Plasmas et des Mat\'eriaux Conducteurs et Leurs
Application, D\'epartement de Physique, Facult\'e des
Science,Universit\'e des Sciences et de la Technologie d'Oran
(U.S.T.O),Oran 31000, Algeria}

{(b) \it D\'epartement de Physique, Facult\'e des Sciences,
Universit\'e des Sciences et de la Technologie d'Oran (U.S.T.O),Oran
31000,Algeria}

\end{center}

\begin{flushleft}
\hrulefill

{\bf Abstract}
\end{flushleft}

Molecular process of polymer collapse was reproduced by isothermal
molecular dynamics simulation. The initial polymer chains were
obtained by mean of random walks in continuum space. Two potential
models were considered to represent short range interactions between
monomers. Various structural properties during the collapse of the
polymer were measured and different collapse pathways were observed.
For low temperatures, we have obtained a spontaneous collapse to
helical states.

\begin{flushleft}
{\it PACS: 31.15.Qg; 87.15.Rn}\\
{\it Keywords:} Molecular dynamics; Polymer collapse pathways;
Helical states

\hrulefill
\end{flushleft}

\footnotetext[1]{Corresponding author. Fax: 00213-41421581; e-mail:
aminesabeur@yahoo.fr}

\begin{flushleft}
{\bf 1. Introduction}
\end{flushleft}

During the last four decades, the collapse of polymer chains has
been extensively investigated \cite{Flory,Stephen,Gennes1}. The key
reason for the interest to polymer collapse is the close
relationship of this process to the protein folding, one of the most
challenging problems in molecular biology. It is well known that in
a good solvent, a polymer chain is extended. However, as one
decrease the solvent quality, the polymer chain adopts a
conformation which tries to minimize solvent-polymer contacts and
the final shape of the chain is compact \cite{Gennes2}. The collapse
of polymers has been studied by many authors recently using both
analytical methods and computer simulations
\cite{Imbert,Michel,Rominszowski}. They all strongly suggest that
the polymer collapse is a multistage process. Schnurr, Mackintosh
and Williams \cite{Pereira,Shnurr,Montesi} treated the case of a
single semiflexible polymer in solution by means of Brownian
Dynamics. They found that the semiflexible polymer collapses to a
torus and the globular ground state is unfavorable. Frish and Verga
\cite{Frish1,Frish2} studied the effect of the solvent on the
collapse pathways of a single flexible homopolymer chain. The
results found by this group show that the collapse of an isolated
homopolymer is dominated by the second order coil to globule
transition. Depending on the quench dept they observed intermediate
sausage regime or a set of pearls along the chain followed by shrink
in polymer size to a final spherical globule. These results agree
with the general scenario for the kinetics of the collapse of a
flexible polymer coil presented by de Gennes \cite{Gennes2}.
Although, fully explicit simulations with solvent particles are
still very expensive numerically, mainly in three dimensions, Polson
et al \cite{Polson1,Polson2,Polson3} have made in a series of works
a molecular dynamics simulation of a heteropolymer immersed in an
explicitly modeled solvent. They studied the effect of hydrophobic,
hydrophilic monomers distribution on the collapse process. A more
recent study \cite{Kikuchi}, investigated the role of hydrodynamics
on the kinetics of polymer collapse transition. This study shows
that hydrodynamics interactions speed up the collapse of the polymer
and enhance cooperative motion of the monomers.

The purpose of the present work is to study the effect of the
temperature and the short range interactions potential model on the
collapse pathways of polymers using molecular dynamics simulation.

Simulating protein folding remains challenging and the question is
how can one reduce computational efforts and correctly predict the
folding kinetics?

The paper is organized as follows. In section (2) we describe the
details of the simulation method used in the present study. In
section (3) we report our results for the structural properties
during the process of the polymer collapse than we conclude with a
discussion of the results.

\begin{flushleft}
{\bf 2. Theory and methods}
\end{flushleft}

\begin{flushleft}
{\it 2.1 Molecular model}
\end{flushleft}

We have used the bead spring model to represent a polymer chain of
$N$ monomers. Non bonded monomers interact via a truncated and
shifted Lennard-Jones potential

\begin{equation}
U_{LJ}(r)=\left\{\begin{array}{cc} {4\varepsilon \left[\left(\sigma
/r\right)^{12} -\left(\sigma /r\right)^{6} \right]+C\left(r_{c}
\right)} & {r\le r_{c} } \\ {0} & {else}
\end{array}\right.
\end{equation}

Where  $C\left(r_{c} \right)$ ensures that the potential vanishes at
the cut-off distance $r_{c}=2.5$. We have used standard, reduced
units in our calculations, where all distances and energies are
expressed in terms of Lennard-Jones parameters $\sigma$ and
$\varepsilon$, respectively; mass is expressed in terms of the
monomer mass $m$ and the unit of time is
$\sqrt{m\sigma^2/\varepsilon}$. Temperature and energy are made
numerically identical by setting the Bolzmann constant $k_B$ to
unity.

We consider two different potential models for representing the
nearest neighbor monomers connectivity, henceforth referred to as
potential models A and B. In potential model A, monomer $i$
interacts with monomer $i-1$ and monomer $i+1$ via a strong
anharmonic potential defined by

\begin{equation}
 U_{A} (r)=a\left(r-r_{0} \right)^{2} +b\left(r-r_{0} \right)^{4}
\end{equation}

Here  $r_{0}=0.85$  is the equilibrium distance. We take $a=30$ and
$b=100$ for the constants characterizing the harmonic and anharmonic
interactions as in \cite{Frish1,Frish2}. In case of the potential
model B, the interactions between pairs of bonded monomers are
represented only by a harmonic potential of the form

\begin{equation}
U_{B} (r)=a\left(r-r_{0} \right)^{2}
\end{equation}

\begin{flushleft}
{\it 2.2 Initial Configuration}
\end{flushleft}

The initial configuration of the polymer chain is a three
dimensional random walk in continuum space . Starting from a
position at the origin $r_1(0,0,0)$, the positions $(x_i,y_i,z_i)$
of the monomers are generated using the following procedure

\begin{equation}
\left\lbrace
\begin{array}{l}
x_i = x_{i-1} + a cos(\phi)sin(\theta)\\
\\
y_i = y_{i-1} + a sin(\phi)sin(\theta)\\
\\
z_i = z_{i-1} + a cos(\theta)\\
\end{array}
\right.
\end{equation}

$\theta$  and $\phi$ are two uniformly random angles chosen
respectively in the ranges $[0,\pi/9]$ and $[0,2\pi]$ and $a$ is the
distance between two successive monomers. $\theta$ has to be small
to simulate a stretched polymer chain. In Figure (1), we show a
snapshot of a polymer chain generated by this procedure.

\begin{flushleft}
{\it 2.3 Polymer Dynamics}
\end{flushleft}

We have performed molecular dynamics simulations in the canonical
ensemble. In order to maintain the temperature constant, we have
coupled the system to the Nos\'e Hoover Thermostat
\cite{Hoover1,Hoover2}.

This scheme derives from a modified Lagrangian formalism that
introduces a new degree of freedom. The equations of motion are then

\bigskip

\begin{equation}
\left\lbrace
\begin{array}{l}
\dot{r_{i}}=p_i/m_i\\
\\
\dot{p_i}=-\frac{\partial V_i}{\partial r_i}-\xi \dot{p_i}\\
\\
\dot{\xi}=\frac{1}{Q}\bigl[\sum_{i=1}^{N}\frac{p_i^2}{m_i}-3Nk_{B}T\bigr]\\
\end{array}
\right.
\end{equation}

\bigskip

Here $r_i$ and $p_i$ are respectively the position and momentum of
monomer $i$. $Q$ is a parameter defining thermal inertia in the
system. it must be carefully chosen to prevent the simulation from
unwanted oscillations. In our simulations, we have chosen $Q=10$.
$K_B$ represents the Boltzman constant and $\xi$ is the friction
variable. We have solved numerically the equations of motion, using
the Verlet Newton-Raphson algorithm \cite{Frenkel} and the time step
was set to $\Delta t=10^{-3}$.

\newpage

\begin{flushleft}
{\it 2.4 Observables}
\end{flushleft}

The quantities of central interest sampled during the simulations
are the gyration radius $R_g$ of the polymer and the long range
order parameter $S$. For a single chain conformation $\{r_N\}$, they
are defined respectively by

\begin{equation}
R_g=\sqrt{\frac{1}{N}\sum_{i=1}^N (r_i-r_{c.m.})^2}
\end{equation}

where $r_i$ is the position of the ith monomer, $N$ is the number of
monomers, and $r_{c.m.}$ is the center of mass position of the
polymer chain given by $r_{c.m.}=\frac{1}{N}\sum_{i=1}^N r_i$ and

\begin{equation}
S = \frac{1}{(N-2)}\bigl|\sum_{k=1}^{N-2}(b_{k}\times b_{k+1})\bigr|
\end{equation}

The parameter $S$ measures the long-range order present in the
folded structures.

\begin{flushleft}
{\it 2.5 Locating the $\Theta$-point}
\end{flushleft}

From theorical point of view, the $\Theta$-point is the temperature
at which the distribution of the end-to-end distance of the polymer
chain $P(R)$ display gaussian statistics

\begin{equation}
P(R)=4\pi R^2 A^\frac{3}{2}exp(-\pi A R^2)
\end{equation}

Where $A=3/(2\pi \langle R^2 \rangle)$ and $R$ is the end-to-end
vector defined by

\begin{equation}
R=\sum_{i=1}^N b_i
\end{equation}

Here, $b_i$ is the bond vector.

\bigskip

We have used a method described by Yong et al \cite{Yong} and based
on Rosenbluth sampling \cite{Grassberger} for locating the
$\Theta$-point. The main idea of this method is to compute the
distribution of the end to end distance and compare it to the
theorical distribution $P(R)$ for different values of the
temperature. In our simulations, we have explored the temperature
range $T=3.0-4.0$ and found the $\Theta$-point at the temperature
$\Theta \approx 3.4$ for the set of parameters used in the molecular
model presented above. We show in figure (2) that for a polymer
chain size $N=256$, the end-to-end distance distribution fit closely
the theorical distribution for the temperature $\Theta \approx 3.4$.

\newpage

\begin{flushleft}
{\bf 3. Results and discussion}
\end{flushleft}

We have investigated the collapse pathways of the polymer chains
after an abrupt decrease in temperature. We first consider the case
of small quench below the $\Theta$-point $T=3.0$. Immediately after
the quench, pearls start to grow along the polymer chain backbone,
these pearls nucleate more easily from the polymer chain extremities
as described in \cite{Frish1}. The high mobility of the monomers
helps the pearls to merge rapidly to a compact globule. Figures (3)
and (4) show a representative picture of this process. It is
important to note that for a small temperature quench, similar
collapse pathways were observed during the simulations either using
the potential models A or B.

In the case of a strong quench $T=0.1$, the polymer collapse pathway
is completely different. We have observed a well ordered helix
structure along the chain. Figure (5) and (6) show the spontaneous
collapse of the polymer chain into this structure. (The polymer is
drawn as a tube to see clearly the helix structure).

The rate of collapse kinetics was measured by monitoring the
variation of the gyration radius $R_g$ with time. Figure (7) shows
that the deeper the temperature quench, the longer the collapse.

We found that in simulations using potential model A, polymer chains
show a better folding behaviors and a more pronounced helical ground
states than in simulations using potential model B, in agreement
with \cite{Clementi}, where it was argued that a plain harmonic
potential could induce energy localization in some specific modes
and significantly increase the time for equilibration. This result
is clearly visible in Figure (8) through the plot of the long range
order parameter $S$ as a function of time for both potential models
A and B.

The numerical simulation of the collapse pathways of polymer chains
in the present work and its comparison with previous studies can be
summarized as follows:

(1) The process of collapse of polymer chains is a two stage
process, in agreement with the results found by Frish and Verga
\cite{Frish1}. For a small temperature quench below the
$\Theta$-point, the first stage is characterized by the formation of
pearls along the chain and in the second stage, the polymer becomes
compact. For a strong temperature quench, the polymer collapses
spontaneously to an ordered helix structure without the use of any
torsional potential as in \cite{Rapaport1,Rapaport2} or a
complicated potential form as in \cite{Kemp}. Special ground states
like double helices with antiparallel alignment have been also
obtained at some temperature quenches (see figure (9)).

(2) The polymer chain collapse seems to be much slower for a
strong temperature quench below the $\Theta$-point.

(3) The use of potential model A shows a better ordered structures
at low temperatures.

The simplified approach of this work has been used in a polymer
context, however it is also relevant to protein folding as the
helical states are the main ground states observed experimentally
\cite{Cao}. A better investigation of those ground states would be a
future possible direction. The role of monomers relative contact
order on the spontaneous polymer collapse into helical states is
under investigation.

\begin{flushleft}
{\bf Acknowledgements}
\end{flushleft}

The authors would like to thank Professor F. Schmid from the
Condensed Matter Group at Bielefeld University for valuable and
fruitful discussions and suggestions.

\begin{figure}[ht]
\begin{center}
\includegraphics{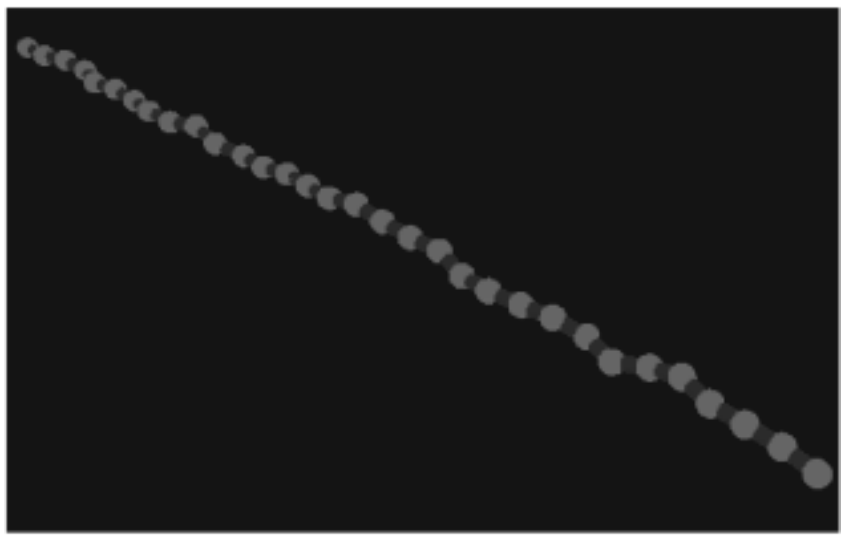}
\caption{Initial configuration of the polymer chain}
\end{center}
\end{figure}

\begin{figure}[ht]
\begin{center}
\includegraphics{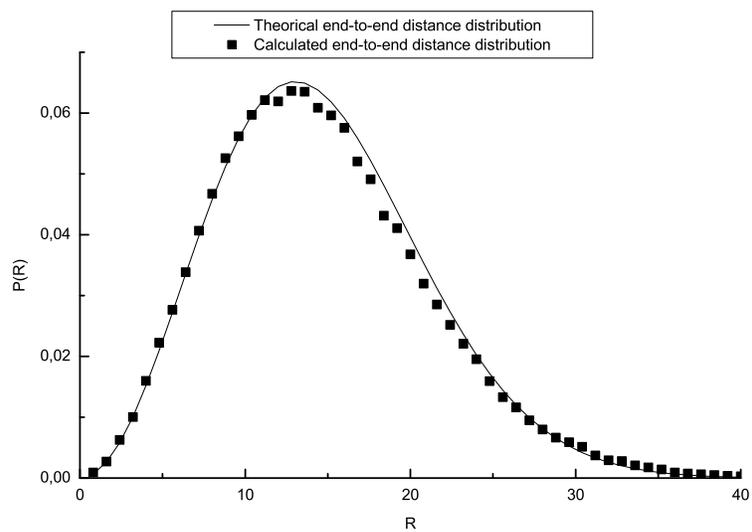}
\caption{End-to-end distance distribution for the temperature
$\Theta=3.4$}
\end{center}
\end{figure}

\begin{figure}[ht]
\begin{center}
\includegraphics{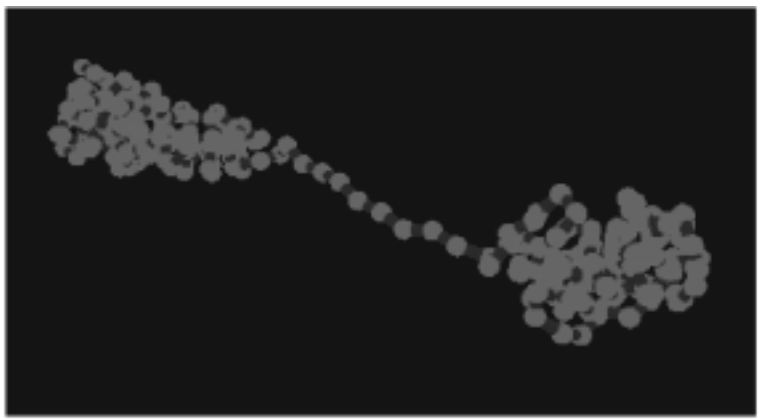}
\caption{Molecular dynamics simulation of a $N=256$ polymer chain at
temperature $T=3.0$, pearls regime at $t=4\times10^4\Delta t$}
\end{center}
\end{figure}

\begin{figure}[ht]
\begin{center}
\includegraphics{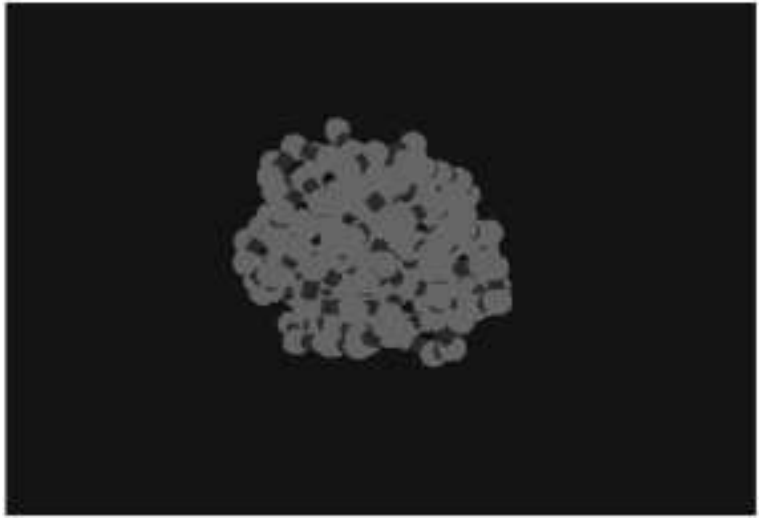}
\caption{Molecular dynamics simulation of a $N=256$ polymer chain at
temperature $T=3.0$, globular state at $t=10^5\Delta t$}
\end{center}
\end{figure}

\begin{figure}[ht]
\begin{center}
\includegraphics{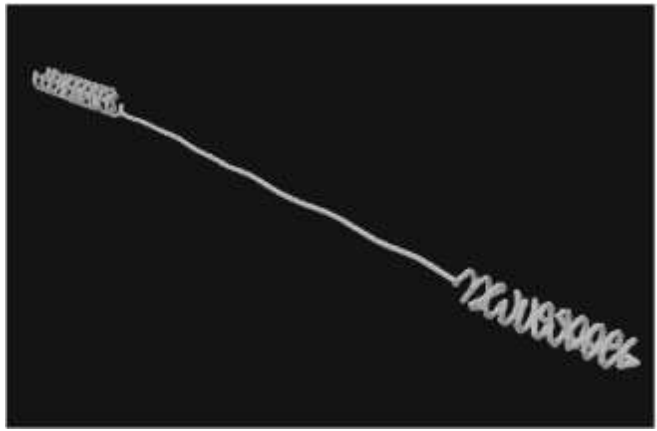}
\caption{Molecular dynamics simulation of a $N=256$ polymer chain at
temperature $T=0.1$, early stages of helix formation at
$t=15\times10^4\Delta t$}
\end{center}
\end{figure}

\begin{figure}[ht]
\begin{center}
\includegraphics{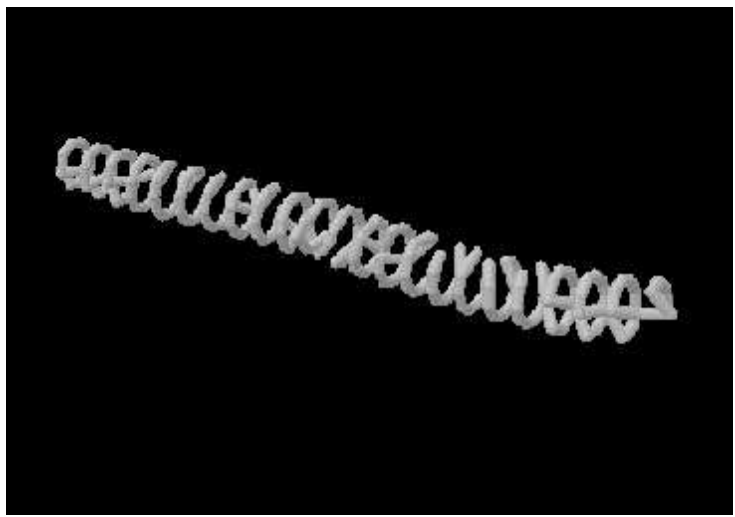}
 \caption{Molecular dynamics simulation of a $N=256$ polymer chain at temperature
$T=0.1$, a well formed helix at $t=3\times10^5\Delta t$}
\end{center}
\end{figure}

\begin{figure}[ht]
\begin{center}
\includegraphics{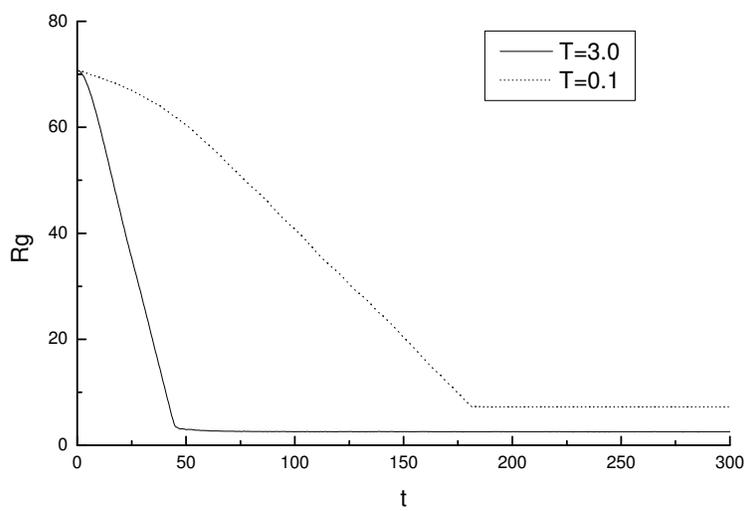}
 \caption{Gyration radius $R_g$ versus time for two differents temperatures
$T=0.1$ and $T=3.0$ and polymer chain size $N=256$}
\end{center}
\end{figure}

\begin{figure}[ht]
\begin{center}
\includegraphics{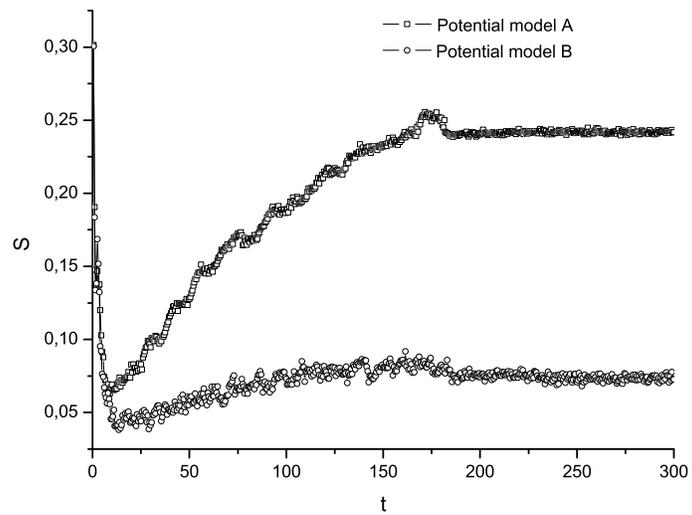}
 \caption{Order parameter as function of time for polymer chain size $N=256$ at temperature $T=0.1$
 using potential model A and potential model B}
\end{center}
\end{figure}

\begin{figure}[ht]
\begin{center}
\includegraphics{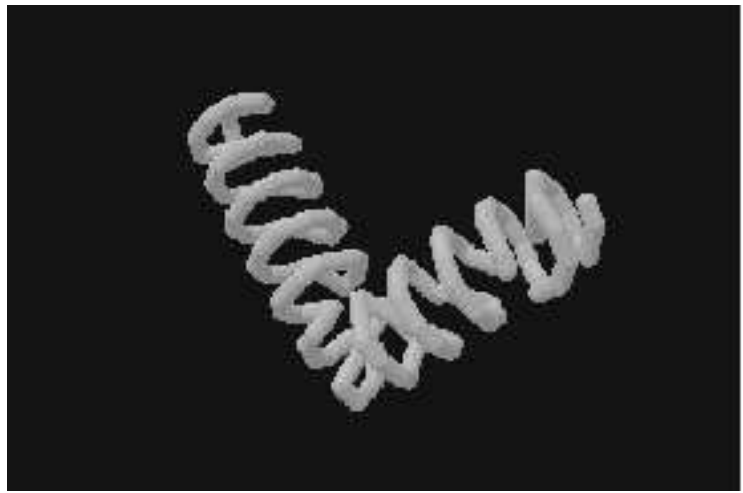}
 \caption{Molecular dynamics simulation of a $N=128$ polymer chain at temperature
$T=0.05$, a pair of helices with antiparallel alignement at $t=15\times10^4\Delta t$}
\end{center}
\end{figure}

\end{document}